\documentclass[reprint, amsmath,amssymb,aip, apl, floatfix]{revtex4-1}
\usepackage[utf8]{inputenc}
\usepackage[english]{babel}
\usepackage{hyperref}
\usepackage{dcolumn}
\usepackage{bm}
\usepackage{hyperref}
\usepackage[dvipsnames]{xcolor}
\usepackage{graphicx}
\usepackage[capitalise]{cleveref}
\usepackage{gensymb}
\usepackage{textcomp}
\usepackage{tabularx, cellspace}
\usepackage{bold-extra} 
\usepackage{bold-extra}
\usepackage{tabularx}

\usepackage[normalem]{ulem}

\definecolor{gregRed}{RGB}{209, 0, 11}

\newcommand{\greg}[1]{\textcolor{gregRed}{#1}}

\newcommand{\gregsout}{\bgroup\markoverwith{\textcolor{gregRed}{\rule[0.5ex]{2pt}{0.4pt}}}\ULon}

\hypersetup{
 pdftitle={},    % title
 pdfauthor={Gregory Moille},     % author
 pdfcreator={Gregory Moille},   % creator of the document
 pdfproducer={LaTeX}, % producer of the document
 pdfkeywords={Version 1.0},
 colorlinks=true,       % false: boxed links; true: colored links
 linkcolor=blue,          % color of internal links (change box color with
 citecolor=blue,        % color of links to bibliography
 filecolor=blue,      % color of file links
 urlcolor=blue}

\newcommand{\beginsupplement}{%
 \setcounter{table}{0}
 \renewcommand{\thetable}{S\arabic{table}}%
 \setcounter{figure}{0}
 \renewcommand{\thefigure}{S\arabic{figure}}%
 \renewcommand{\thesection}{S-\arabic{section}}
}
\usepackage{titlesec}
\titleformat*{\section}{\bfseries\Large}
\titleformat*{\subsection}{\bfseries}
\usepackage{caption}

\begin{document}
%       .o.                       .   oooo                                    
%      .888.                    .o8   `888                                    
%     .8"888.     oooo  oooo  .o888oo  888 .oo.    .ooooo.  oooo d8b  .oooo.o 
%    .8' `888.    `888  `888    888    888P"Y88b  d88' `88b `888""8P d88(  "8 
%   .88ooo8888.    888   888    888    888   888  888   888  888     `"Y88b.  
%  .8'     `888.   888   888    888 .  888   888  888   888  888     o.  )88b 
% o88o     o8888o  `V88V"V8P'   "888" o888o o888o `Y8bod8P' d888b    8""888P' 
\author{Gr\'egory~Moille}
\email{gmoille@umd.edu}
\affiliation{Joint Quantum Institute, NIST/University of Maryland, College Park, USA}
\affiliation{Microsystems and Nanotechnology Division, National Institute of Standards and Technology, Gaithersburg, USA}
\author{Daron~Westly}
\affiliation{Microsystems and Nanotechnology Division, National Institute of Standards and Technology, Gaithersburg, USA}
\author{Edgar~F.~Perez}
\affiliation{Joint Quantum Institute, NIST/University of Maryland, College Park, USA}
\affiliation{Microsystems and Nanotechnology Division, National Institute of Standards and Technology, Gaithersburg, USA}
\author{Meredith~Metzler}
\affiliation{Center for Nanoscale Science and Technology, National Institute of Standards and Technology, Gaithersburg, USA}
\author{Gregory~Simelgor}
\affiliation{Microsystems and Nanotechnology Division, National Institute of Standards and Technology, Gaithersburg, USA}
\author{Kartik Srinivasan}
\affiliation{Joint Quantum Institute, NIST/University of Maryland, College Park, USA}
\affiliation{Microsystems and Nanotechnology Division, National Institute of Standards and Technology, Gaithersburg, USA}
\date{\today}

%  ooooooooooooo  o8o      .   oooo            
% 8'   888   `8  `"'    .o8   `888            
%      888      oooo  .o888oo  888   .ooooo.  
%      888      `888    888    888  d88' `88b 
%      888       888    888    888  888ooo888 
%      888       888    888 .  888  888    .o 
%     o888o     o888o   "888" o888o `Y8bod8P' 
\title{Integrated Buried Heaters for Efficient Spectral Control of Air-Clad Microresonator Frequency Combs}

%       .o.        .o8                    .                                    .   
%      .888.      "888                  .o8                                  .o8   
%     .8"888.      888oooo.   .oooo.o .o888oo oooo d8b  .oooo.    .ooooo.  .o888oo 
%    .8' `888.     d88' `88b d88(  "8   888   `888""8P `P  )88b  d88' `"Y8   888   
%   .88ooo8888.    888   888 `"Y88b.    888    888      .oP"888  888         888   
%  .8'     `888.   888   888 o.  )88b   888 .  888     d8(  888  888   .o8   888 . 
% o88o     o8888o  `Y8bod8P' 8""888P'   "888" d888b    `Y888""8o `Y8bod8P'   "888" 
\begin{abstract}
    \noindent Integrated heaters are a basic ingredient within the photonics toolbox, in particular for microresonator frequency tuning through the thermo-refractive effect. Resonators that are fully embedded in a solid cladding (typically SiO\textsubscript{2}) allow for straightforward lossless integration of heater elements. However, air-clad resonators, which are of great interest for short wavelength dispersion engineering and direct interfacing with atomic/molecular systems, do not usually have similarly low loss and efficient integrated heater integration through standard fabrication. Here, we develop a new approach in which the integrated heater is embedded in SiO$_2$ below the waveguiding layer, enabling more efficient heating and more arbitrary routing of the heater traces than possible in a lateral configuration. We incorporate these buried heaters within a stoichiometric Si$_3$N$_4$ process flow that includes high-temperature ($>$1000~$^\circ$C) annealing. Microring resonators with a 1~THz free spectral range and quality factors near 10$^6$ are demonstrated, and the resonant modes are tuned by nearly 1.5~THz, a 5$\times$ improvement compared to equivalent devices with lateral heaters\greg{.} Finally, we demonstrate broadband dissipative Kerr soliton generation in this platform, and show how the heaters can be utilized to aid in bringing relevant lock frequencies within a detectable range.

\end{abstract}

\maketitle

% ooo        ooooo            o8o              
% `88.       .888'            `"'              
%  888b     d'888   .oooo.   oooo  ooo. .oo.   
%  8 Y88. .P  888  `P  )88b  `888  `888P"Y88b  
%  8  `888'   888   .oP"888   888   888   888  
%  8    Y     888  d8(  888   888   888   888  
% o8o        o888o `Y888""8o o888o o888o o888o 

\begin{figure*}
    \includegraphics[width=\textwidth]{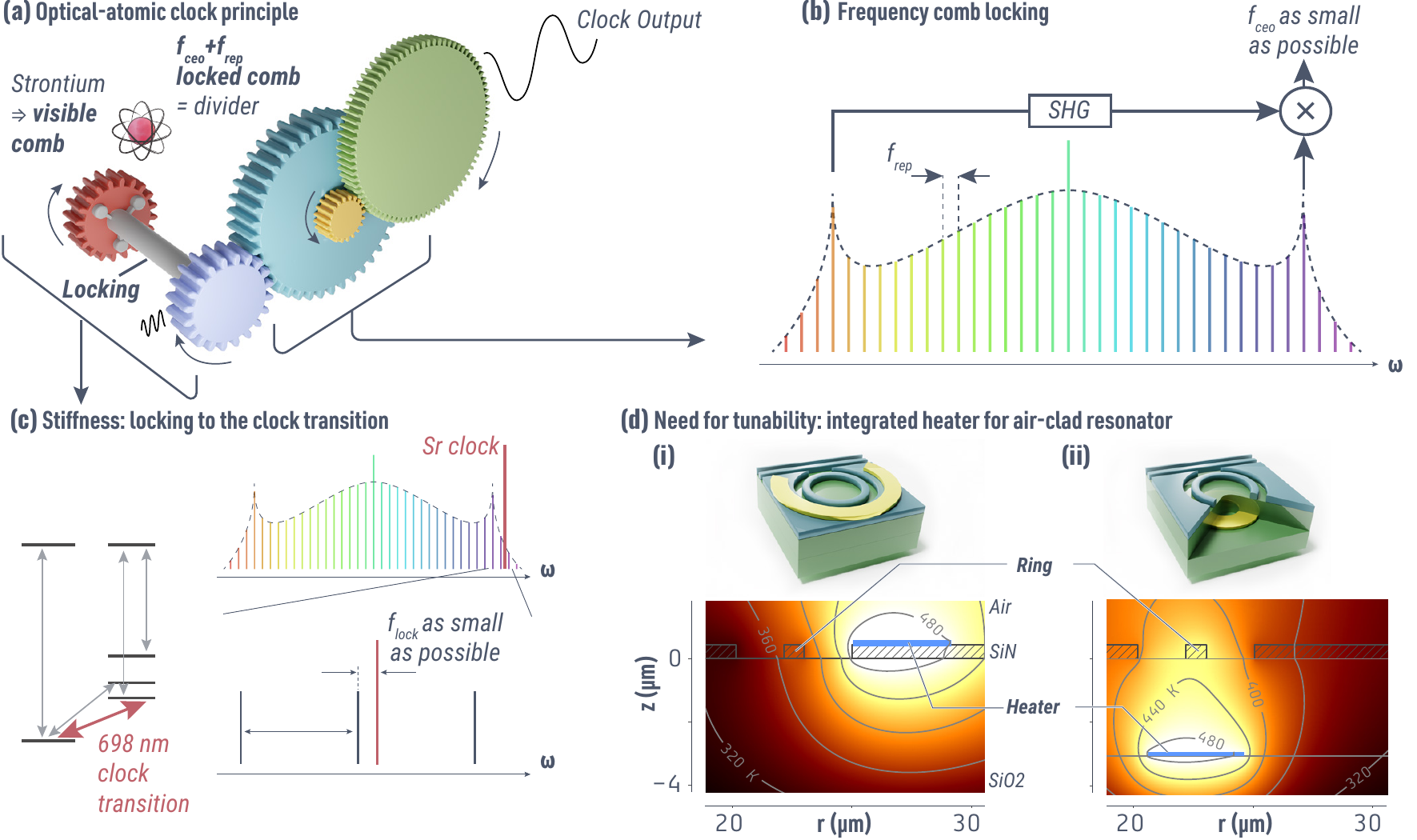}
    \caption{\label{fig:1} 
    \textbf{Optical atomic clock principle and stiffness of the frequency comb locking frequency -- } %
    \textbf{(a)} Concept of an optical frequency comb acting as a frequency divider, depicted as a gear box and inspired by Ref.~\cite{DiddamsScience2020} where the stability of a comb tooth in the optical domain is divided down to the microwave domain with the clock output being the microcomb repetition rate. For the division to be as low noise as possible, the frequency comb needs to be locked. This means that the carrier envelope offset ($f_\mathrm{ceo}$) and a single comb tooth frequency need to be locked. \textbf{(b)} Concept of $f_\mathrm{ceo}$ detection and locking. A tooth from an octave-spanning frequency comb and the second harmonic of a low frequency tooth from the same comb are interfered with each other to retrieve $f_\mathrm{ceo}$, which can in principle range from 0 to $f_\mathrm{rep}/2$ and thus potentially be outside the photodetector bandwidth when $f_\mathrm{rep}$ is large (typical for microcombs). The octave span of the microcomb is obtained thanks to dispersive waves (DWs), where the cavity dispersion allows for comb tooth resonance enhancement at the targeted frequencies. \textbf{(c)} The clock relies on the locking of one comb tooth to the optical transition of an atomic reference, for instance  \textsuperscript{85}Sr at 698~nm. Therefore, one comb tooth must be close enough to this optical transition to be able to be locked. As for $f_\mathrm{ceo}$, this locking frequency $f_\mathrm{lock}$ can be between 0 and $f_\mathrm{rep}/2$. \textbf{(d)} As dispersion mostly impacts the position of a comb tooth that is amplified through DW enhancement, a control knob to shift the whole frequency comb is needed. Integrated heaters that leverage the thermo-refractive effect can provide this functionality. However, an air-clad ring is needed to support the dispersion required to reach the 698~nm wavelength of the \textsuperscript{85}Sr reference. Ergo, two possibilities exist: lateral heating (i) and the new buried heater technology introduced here. (ii). Finite element method simulations show that the heat build-up is much more efficient in the case of the buried heater, with a temperature inside the ring reaching 420~K against only 365~K for the same heating power. }
\end{figure*}

Since their discovery, frequency combs have yielded a plethora of applications -- spectroscopy~\cite{DuttSci.Adv.2018}, optical clocks~\cite{NewmanOptica2019}, frequency synthesis~\cite{SpencerNature2018}, and distance ranging~\cite{SuhScience2018}. %
Their integration on chip through the use of dissipative Kerr soliton (DKS) states in different platforms~\cite{LeoNaturePhoton2010, HerrNaturePhoton2014, XueNaturePhoton2015, LiOptica2017,GuoNaturePhys2017, MoilleLasersPhotonicsRev.2020} has led to a focus on low power~\cite{SternNature2018} and low footprint~\cite{YeLaserPhotonicsRev.2022} devices for deployable metrology outside of the laboratory~\cite{DiddamsScience2020}. %
The realization of octave-spanning microcombs has been made possible by harnessing unprecedented control of integrated microresonators dispersion~\cite{DiddamsScience2020}, that additionally overlap with the atomic optical transition frequency~\cite{YuPhys.Rev.Applied2019} has helped spur interest in their use in portable optical atomic clocks~\cite{DrakePhys.Rev.X2019,NewmanOptica2019}. %
In such applications, the frequency comb acts as a gear box~\cite{DiddamsScience2020} translating the optical frequency stability of a comb tooth locked to an atomic transition to a microwave frequency through the DKS repetition rate [\cref{fig:1}(a)]. However, the comb needs to be fully stabilized {when realizing such a} frequency division scheme. In the clock case, the carrier envelope offset $f_\mathrm{ceo}$ (i.e. the shift from the zero frequency) [\cref{fig:1}(b)] needs to be locked along with a comb tooth close enough from the optical atomic transition frequency, here called $f_\mathrm{lock}$ [\cref{fig:1}(c)].% 
Their locking makes the system in~\cref{fig:1}(a) stiff, in the sense that only a single set of geometric parameters (ring width and thickness combination) will provide for enough power enhancement at the frequencies of interest (often at the location of dispersive waves (DWs)) while bringing a comb tooth sufficiently close to the atomic transition frequency. %
Yet, each of these goals are essentially driven by two different parameters: the DW spectral position is defined by the cavity dispersion while $f_\mathrm{lock}$ (the beat note between a comb tooth and the optical atomic transition) and $f_\mathrm{ceo}$ can be controlled by a simple uniform spectral shift of the comb. %
Therefore, integrated heaters appear as a suitable solution, leveraging spectral tuning via the thermo-refractive effect~\cite{BosmanPhys.Rev.1963} and compatibility with $\chi^{(3)}$ microcomb platforms. Frequency tuning ranging up to hundreds of gigahertz has been demonstrated~\cite{XueOpt.ExpressOE2016, JoshiOpt.Lett.OL2016, HelgasonNat.Photonics2021, TikanNat.Phys.2021}, though mostly in resonators fully embedded in a silica cladding.

The use of air-clad devices is, however, desirable for dispersion engineering to reach atomic transition frequencies in the short near-infrared and near-visible~\cite{MoilleCLEOSci.Innov.2022} and allows for post-fabrication processing to tune the geometrical dispersion if needed~\cite{MoilleAppl.Phys.Lett.2021}. Direct integration of the heater to the side of the ring results in poor heat build-up at the ring core [\cref{fig:1}(d.i)], caused by the air trenches that act as an insulator but are essential to create the ring during the lithography and etch fabrication steps. 
The alternative approach we propose in this paper is to bury the integrated heater below the ring resonator [Fig.~\ref{fig:1}(d,ii)], where a few micrometer gap of silica separates the metal from the optical layer, resulting in no change in the ring losses. Due to the continuous path for heat transfer between the heater and ring layers, the efficiency of such a buried heater is much higher than for the lateral one, confirmed by thermal simulation [\cref{fig:1}(d)]. We demonstrate the fabrication of such unique integrated heaters, which are wire-bonding ready for integration in optical clock systems. In addition, fully embedding the heater in silica and subsequently planarizing the silica layer makes the remainder of the fabrication process compatible with a typical Si photonics process flow ~\cite{ThomsonJ.Opt.2016}. We experimentally show the high tuning efficiency of the system, where a resonator mode shift of 1.5~THz (equal to 1.5$\times$ the resonator free spectral range (FSR)) is achieved without degradation of the resonator quality factor and with limited cross-talk. While tuning of air-clad resonators has numerous potential applications, including in sensing and coupling to quantum emitters, we showcase the utility of these heaters in the context of broadband DKS microcombs. We experimentally demonstrate tunability of the microcomb with an estimated shift of $f_\mathrm{ceo}$ of 15~\% of the FSR and an estimated shift of the locked optical clock frequency $f_\mathrm{lock}$ of close to 2~FSR, while little modification to dispersion -- and hence the DW position -- occurs.

\begin{figure}[!t]
    \includegraphics[width=0.95\columnwidth]{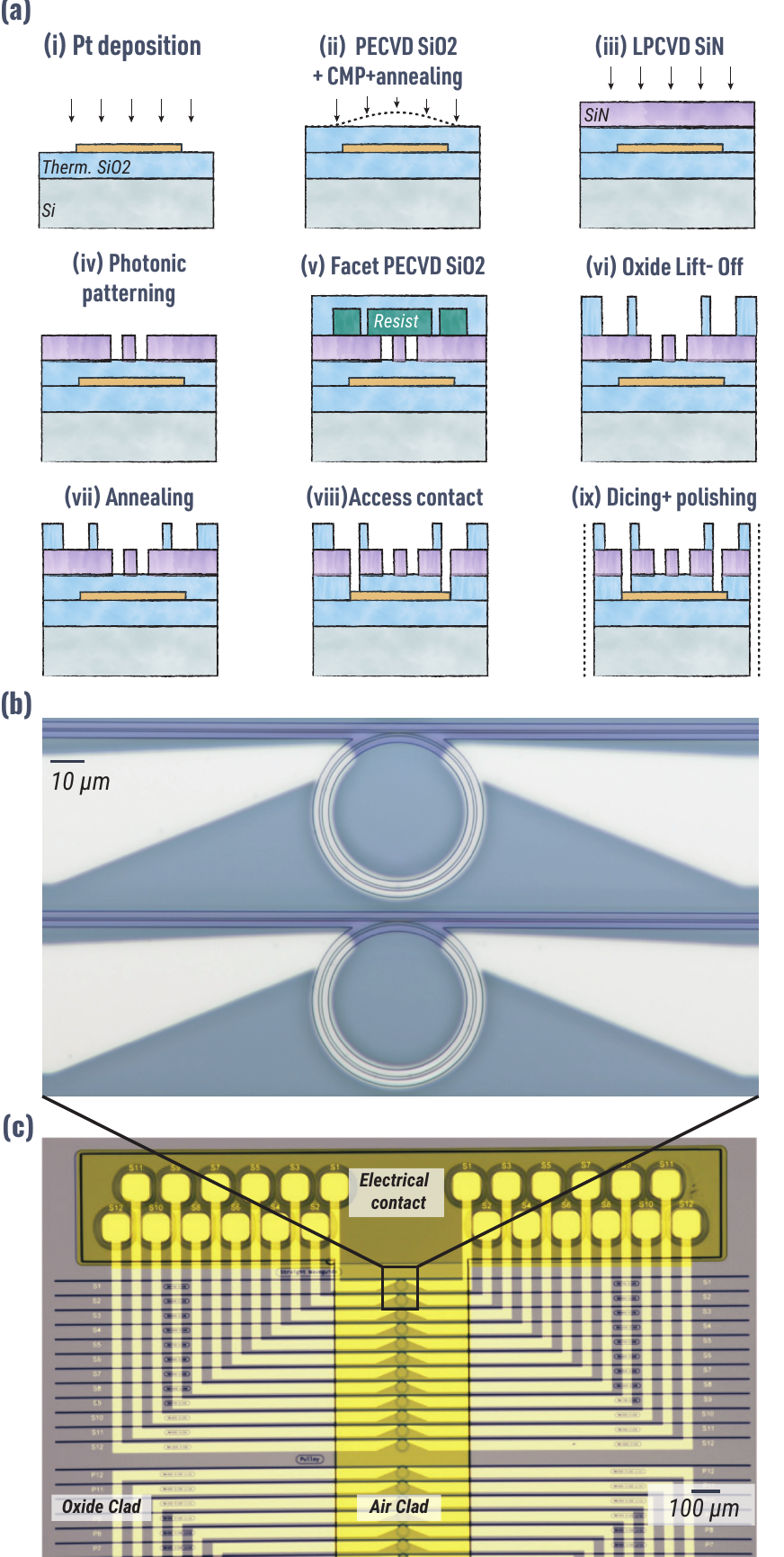}
    \caption{\label{fig:2} %
    \textbf{Buried heater fabrication --} %
    \textbf{(a)} Process flow starting with a commercial silicon wafer with thermal silicon dioxide (SiO\textsubscript{2}). (i) Deposition of the platinum heater layer, lifted-off from previously patterned resist. (ii) Burying the heater in another layer of (HDPCVD) SiO\textsubscript{2}. This layer needs to be planarized through CMP because of the metal underneath. An annealing step at 1000~{\degree}C for 3 hours is then performed. (iii) Resumption of the standard fabrication process with 430~nm Si\textsubscript{3}N\textsubscript{4} film deposition through LPCVD followed by (iv) e-beam patterning and RIE etching to create the photonics layer. (v-vii) Oxide lift-off using low temperature HDPCVD for air-clad rings with oxide-clad facets, followed by annealing. (viii) A last etch step through Si\textsubscript{3}N\textsubscript{4} and SiO\textsubscript{2} to enable electrical access to the heater layer. (ix) Dicing and polishing of the chip for testing. \textbf{(b)} Optical image of the buried heater to demonstrate the alignment of the metal layer with the photonics one. \textbf{(c)} Optical image of the chip, showing how the buried heater fabrication enables arbitrary routing of the metal layer with respect to the waveguide layer.}
\end{figure}

\begin{figure}[!t]
    \includegraphics[width=\columnwidth]{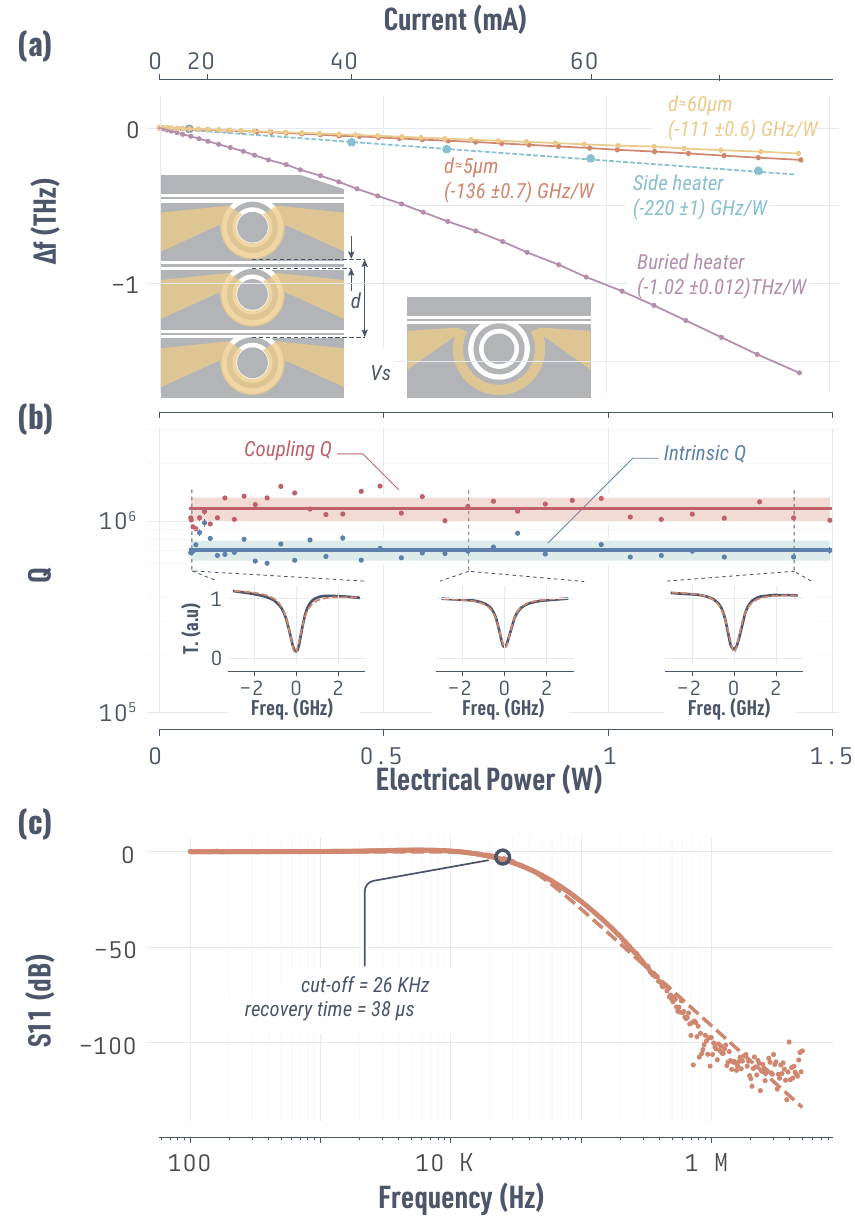}
    \caption{\label{fig:3} %
    \textbf{Linear characterization of the buried heater resonator system --} %
    \textbf{(a)} Resonant frequency shift probed around 309~THz (970~nm) for the buried heater compared to the lateral heater for the same nominal resonator-waveguide system. Cross-talk is characterized by measuring the frequency shift of neighboring rings, resulting in a cross-talk of $\Xi \approx 13~\%$ for a neighboring device separated by 5~$\mu$m from the target device. Devices with a larger separation show only a weak reduction in $\Xi$. %
    \textbf{(b)} Intrinsic ($Q_\mathrm{0}$, blue) and coupled ($Q_\mathrm{c}$, red) quality factor for different electrical power, \textit{i.e.} temperature. The absence of  variation in $Q_\mathrm{c}$ shows that no noticeable thermal expansion happens in the system, and the geometry remains overall unchanged. The error bar is within the marker size. The three insets are representative transmission spectra at different power (temperature). %
    \textbf{(c)} Frequency response of the buried heater under small-signal modulation, using a vector network analyzer to apply the modulation and coherently photodetect the transmitted signal. The value of $\approx$~38~{\textmu}s obtained at the 3~dB cut-off is {consistent} with the typical bandwidths of state-of-the-art thermal frequency shifters.
    }
\end{figure}

% /*oooooooooooo            .o8       
% `888'     `8           "888       
%  888          .oooo.    888oooo.  
%  888oooo8    `P  )88b   d88' `88b 
%  888    "     .oP"888   888   888 
%  888         d8(  888   888   888 
% o888o        `Y888""8o  `Y8bod8P' 
                                  
The fabrication of the system [\cref{fig:2}(a)] starts with a standard commercial silicon substrate with a 3~{\textmu}m thick thermal oxide. We then define the heater pattern using a direct-write lithography maskless aligner (MLA)~\cite{NISTdisclaimer} [\cref{fig:2}(a-i)]. The platinum is then deposited (electorn beam evaporation) and lifted off. Another layer of 2.8~{\textmu}m of SiO\textsubscript{2} is deposited above the heater layer using a 180~{\degree}C low temperature High Density Plasma Chemical Vapor Deposition (HDPCVD) process [\cref{fig:2}(a-ii)]. To allow for the best material and the lowest absorption, we proceed to anneal the wafer at 1000~{\degree}C for three hours. Thanks to the low diffusivity of metal into silica~\cite{McBrayerJ.Electrochem.Soc.1986}, the buried heater remains geometrically intact and the silica layer exhibits low absoprtion after the annealing. A chemical-mechanical polishing (CMP) step [\cref{fig:2}(a-ii)] is done to create a flat surface before low-pressure chemical vapor deposition (LPCVD) of a 430~nm thick film of silicon nitride~\cite{MoilleOpt.Lett.OL2021} [\cref{fig:2}(a-iii)]. From this point on, the process is the same as our regular nano-patterning and fabrication of air-clad ring resonators~\cite{MoilleAppl.Phys.Lett.2021}. The ring resonator patterns are created through electron-beam lithography, with the patterns aligned to the underlying buried heaters using metal alignment marks that are created within the heater layer. The electron-beam resist is used as a mask for a CHF\textsubscript{3} reactive ion etch (RIE) of the silicon nitride [\cref{fig:2}(a-iv)]. In order to maintain an air-clad ring resonator for dispersion purposes while allowing for low insertion losses, a lift-off of low temperature HDPCVD SiO\textsubscript{2} using nLOF resist is performed [\cref{fig:2}(a-v,vi)]. A last annealing step is performed to anneal the silicon nitride film, which has been shown to reduce absorption in the C-band through reducing N-H bonds~\cite{SteinThinSolidFilms1983} [\cref{fig:2}(a-vii)]. Finally, an inductively coupled plasma (ICP) etch through the silicon nitride and the substrate silica, combined with a one minute 6:1 buffer oxide etch dip,  reveals the metal layer for electrical contact [\cref{fig:2}(a-viii)]. The chip is then diced and polished for optical testing [\cref{fig:2}(a-ix)].

\begin{figure*}[t]
    \includegraphics[width=\textwidth]{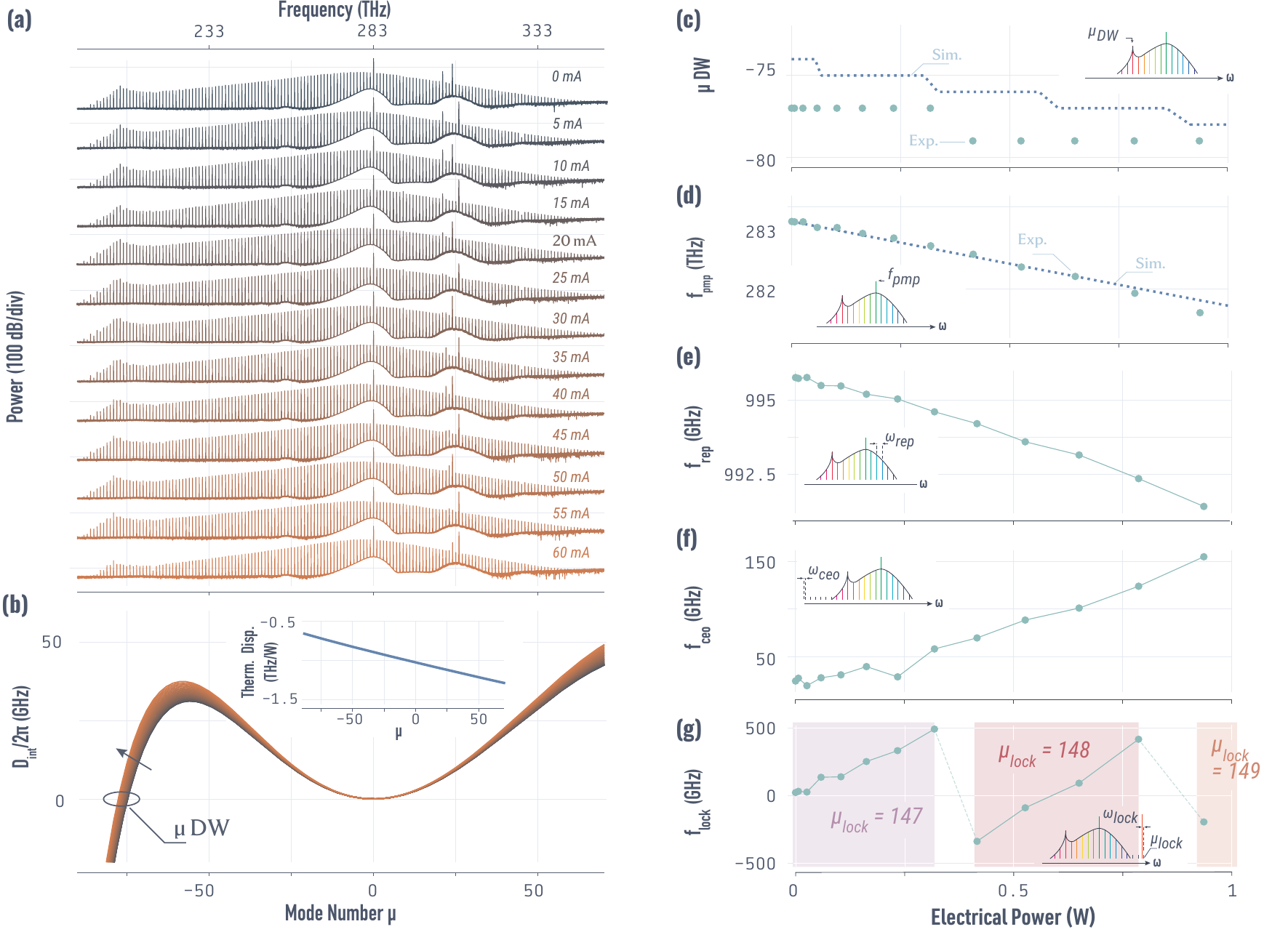}
    \caption{\label{fig:4} \textbf{Tuning and control of the DKS microcomb with the buried heater ---} %
    \textbf{(a)} Single DKS microcomb obtained at different currents (and temperatures) through the buried heater, when pumped around 284~THz (1056~nm). The spectra is obtained through active ``cooling'' of the resonator using a  309~THz to 312~THz cross-polarized and counter-propagating secondary pump. The cooler frequency is changed to obtained the soliton yet its frequency does not impact nonlinear dynamics. The DW at short frequency (long wavelength) is only slightly shifted by 2 comb teeth, showcasing the small, yet not negligible, dispersion of the effective (i.e., mode-weighted) thermo-refractive index. %
    \textbf{(b)} Simulated integrated dispersion for the different current values in (a), accounting for the dispersion in thermo-refractive index (inset) resulting from the variation in modal overlap with the Si$_3$N$_4$, SiO$_2$, and air regions. %
    \textbf{(c)} Shift in dispersive wave mode number with increase in current (electrical power), including both the measured (circles) and simulated (dashed line) values, where the simualtions are based on the integrated dispersion calculation in (b).  %
    \textbf{(d)} Frequency shift of the pump with increase in current (electrical power). %
    \textbf{(e)} Extracted shift of the repetition rate. The repetition rate is obtained through wavemeter calibration of the pump wavelength and measuring with an optical spectral analyzer the comb tooth at $\mu=-76$. The OSA resolution is 20~pm (\textit(i.e.) 2.5GHz resolution), resulting in a measurement error of about 32~MHz (error bars are within the data point size). %
    \textbf{(f)} Extracted $f_\mathrm{ceo}$ shift with electrical power from the extracted $f_\mathrm{rep}$ and the wavemeter calibrated $f_\mathrm{pmp}$. %
    \textbf{(g)} Extracted $f_\mathrm{lock}$ (blue circles) assuming a comb tooth near the \textsuperscript{85}Sr transition at 698~nm. Here, we extrapolate the closest comb tooth frequency based on $f_\mathrm{rep}$ and $f_\mathrm{ceo}$ from (e) and (f). The buried heater allows for tuning of two consecutive comb teeth across the reference frequency (green circles), allowing a full tuning of $f_\mathrm{lock}$ between [-$f_\mathrm{rep}/2$,$f_\mathrm{rep}/2$].
    }
\end{figure*}

% /*ooooo         o8o                                           
% `888'         `"'                                           
%  888         oooo  ooo. .oo.    .ooooo.   .oooo.   oooo d8b 
%  888         `888  `888P"Y88b  d88' `88b `P  )88b  `888""8P 
%  888          888   888   888  888ooo888  .oP"888   888     
%  888       o  888   888   888  888    .o d8(  888   888     
% o888ooooood8 o888o o888o o888o `Y8bod8P' `Y888""8o d888b    

We proceed to characterize the linear performance -- spectral tuning, cross talk and speed -- of the buried integrated heater. The experimental setup consists of a continuous tunable laser (CTL) around 970~nm probing a single resonant mode of the cavity. The chip is placed on an aluminum sample holder not thermally connected to the optical table (see supplementary material). Heaters are addressed from the electrical pads with DC probes connected to a current source, which can produce output up to 1.5W of electrical power. We measure the electrical resistance of the integrated heater to be $R\approx260$~$\Omega$, of the same order of magnitude as typical integrated heaters~\cite{XueOpt.ExpressOE2016, JoshiOpt.Lett.OL2016}. Injecting 1.2~W of electrical power (\textit{i.e.} a current of 67~mA), we measure a frequency shift of the resonance of more than 1.5~THz, corresponding to an $\approx1.5~$FSR red-shift [\cref{fig:3}(a)]; amongst other things, this guarantees the ability to tune one of the microresonator's modes onto resonance with any fixed pump laser wavelength. From simulations (see supplementary materials) this frequency shift is consistent with an increase in temperature at the ring above 900~K. We note that this frequency shift is quite large compared to state-of-the art integrated heaters in the silicon nitride nonlinear photonics platform that can be obtained through foundry fabrication~\cite{RaoLightSci.Appl.2021} where a shift of only $1/2$ of an FSR was achieved for the same electrical power. The efficiency of the buried heater is $\eta_0 = (-1.02\pm0.012)$~THz/W, an improvement by a factor 5 compared to the lateral heater efficiency of $\eta_\mathrm{side} = (-220\pm1)$~GHz/W. The uncertainty estimation arises from the standard deviation in the confidence interval for the linear fit of the frequency shift with electrical power. In addition, the cross talk of the integrated heater is limited. The next ring resonator, with its closest point from the neighboring active heater of about 5~{\textmu}m -- a relatively tight separation considering the waveguide access needed to optically address the resonator -- only show a thermal shift of  $\eta_{+1} = (-136\pm0.7)$~GHz/W. Therefore, the cross-talk of our system can simply be defined by $\Xi = \eta_{+1}/\eta_0 \approx 13~$\%. Interestingly, the cross talk is not highly dependent with the distance from the heater. Indeed, the second neighbor from the active heater shows a spectral shift of $\eta_{+2} = (-111\pm0.6)$~GHz/W, showing that a significant portion of the chip is heating. Therefore, cross-talk could be further improved through an appropriate thermal shunt through the backside of the chip. 

We proceed to consider the impact of the buried heater on the optical quality factor of the ring resonator [\cref{fig:3}(b)]. We found that the intrinsic quality factor, in addition to being similar to that of our regular fabrication without the metal layer of $Q_0\approx 0.75\times 10^6$~\cite{MoilleAppl.Phys.Lett.2021}, remains unchanged for any temperature and electrical power applied to the system. Moreover, we measure that the coupling quality factor $Q_\mathrm{c}~\approx1\times 10^6$ stays fixed at any temperature, highlighting the little thermal expansion of the system, resulting in a fixed geometry with temperature. Our coupling is based on a straight waveguide, but in the future, using pulley-like coupling, which is more sensitive to geometrical variations~\cite{MoilleOpt.Lett.2019}, would give a better estimate to what extent the thermal expansion is truly negligible. %Yet, for the rest of this work, we will consider it negligible. 
We also performed dynamical measurements, in the small signal regime using a vector network analyzer, to measure the recovery time of the buried heater and the frequency up to which the system can function [\cref{fig:3}(c)]. It exhibits a 3~dB cut-off at about 26~kHz, which corresponds to a recovery time of the system of $\approx$38~{\textmu}s. It is worth noting that this value is similar to that of integrated heaters on top of silica-clad rings~\cite{XueOpt.ExpressOE2016}, and is essentially limited by the thermal lifetimes of silicon nitride and silica~\cite{LiOptica2017}.

% ooooo      ooo ooooo          .oooooo.   
% `888b.     `8' `888'         d8P'  `Y8b  
%  8 `88b.    8   888         888      888 
%  8   `88b.  8   888         888      888 
%  8     `88b.8   888         888      888 
%  8       `888   888       o `88b    d88' 
% o8o        `8  o888ooooood8  `Y8bood8P'  

The integrated buried heater presented here shows a high efficiency with a large thermo-refractive mediated spectral shift of the resonant mode. It remains to characterize how the thermal tuning of a ring resonator impact the metric associated with the optical atomic clock frequency comb , namely $f_\mathrm{rep}$, $f_\mathrm{ceo}$ and $f_\mathrm{lock}$. In previous work, it has been demonstrated that thermal tuning of the ring resonator allows for small tuning of the the microcomb repetition rate~\cite{XueOpt.ExpressOE2016}. Here, we also proceed to characterize others metrics, to assess the impact of the thermal tuning on the cavity dispersion, the carrier envelop offset and the locking frequency, which represent three other important quantities for efficient microcomb application in optical atomic clocks.  To do so, we create dissipative Kerr soliton microcombs at different electric powers [\cref{fig:4}(a)]. Interestingly, the overall shape of the frequency comb remains mostly the same, with only a slight shift of the DW position, despite the introduction of another dispersion contribution. Indeed, the variation of the effective refractive index with temperature is wavelength-dependent given the discrepancy in the thermo-refractive coefficient between Si\textsubscript{3}N\textsubscript{4} ($\partial n/\partial T = 2.5\times10^{-5}$~K\textsuperscript{-1})~\cite{ElshaariIEEEPhotonicsJ.2016, MoillePhys.Rev.Appl.2019}, SiO\textsubscript{2} ($\partial n/\partial T = 1\times10^{-5}$~K\textsuperscript{-1})~\cite{ElshaariIEEEPhotonicsJ.2016} and air ($\partial n/\partial T \equiv 0$~K\textsuperscript{-1}) and the variation of the effective mode area, which decreases with higher frequency (or mode number) [\cref{fig:4}(b)]. The integrated dispersion, as obtained from finite element method simulations taking into account Joule heating, temperature distribution, and the thermo-refractive coefficient (see supplementary materials), shows only a slight modification across the thermal tuning range. Therefore, the dispersive waves -- essential for achieving octave spans and self-referencing of a microcomb -- would remain mostly unchanged despite variation of the above parameters. As reported in \cref{fig:4}(c), the experimental DW shift is only 2 modes, from $\mu=-77$ to $\mu=-79$, and shows a slight discrepancy from finite element method simulations ($\mu=-74$ to $\mu=-78$). Although the overall shape of the DKS microcomb remains unchanged, its fundamental characteristics of relevance for clock applications -- repetition rate and carrier-envelope offset frequency -- are tuned with the electrical power injected. As seen previously, the resonances are red-shifted with the electrical power, ergo the pump frequency is also red-shifted [\cref{fig:4}(d)]. As the refractive index of the material increases with the temperature, the group index should also increase, resulting in a lower repetition rate of the DKS circulating in the resonator, as measured experimentally [\cref{fig:4}(e)]. Given the repetition rate shift and the shift of the pump resonance, it is expected that $f_\mathrm{ceo}$ will also vary. We extract it from the measurements of $f_\mathrm{pmp}$ and $f_\mathrm{rep}$ as $f_\mathrm{ceo}=f_\mathrm{pmp} - Nf_\mathrm{rep}$, where $N$ is the integer that brings $f_\mathrm{ceo}$ within the range [-$f_\mathrm{rep}/2$,+$f_\mathrm{rep}/2$], and find that $f_\mathrm{ceo}$ tuning of more than 100~GHz (\textit{i.e} about 10\% of $f_\mathrm{rep}$) is achievable [\cref{fig:4}(f)]. This tuning can be crucial to allow for $f_\mathrm{ceo}$ to be shifted into a detection bandwidth of a photodetector. Finally, we demonstrate the possibility of tuning $f_\mathrm{lock}$ to a stabilized optical reference [\cref{fig:4}(g)]. Here, we have assumed a comb tooth is present near the \textsuperscript{85}Sr transition at 698~nm. In practice, our demonstrated combs in this work do not extend to this wavelength due to poor comb extraction resulting from the straight waveguide coupling scheme employed~\cite{MoilleOpt.Lett.2019}; however, we have demonstrated similar combs (without heaters) extending to 698~nm using optimized pulley couplers~\cite{MoilleCLEOSci.Innov.2022}. Finally, we note that because the $f_\mathrm{lock}$ comb tooth mode order is large, it provides a large multiplicative factor on the temperature-induced shift in $f_\mathrm{rep}$. Thus, with only 1~W of electrical power, $f_\mathrm{lock}$ {shifts across} 3 different comb teeth, constituting a span $>2.5f_\mathrm{rep}$. 

%    .oooooo.   oooo            o8o                        
%  d8P'  `Y8b  `888            `"'                        
% 888           888   .oooo.o oooo   .ooooo.  ooo. .oo.   
% 888           888  d88(  "8 `888  d88' `88b `888P"Y88b  
% 888           888  `"Y88b.   888  888   888  888   888  
% `88b    ooo   888  o.  )88b  888  888   888  888   888  
%  `Y8bood8P'  o888o 8""888P' o888o `Y8bod8P' o888o o888o 
\vspace{2ex}
We have demonstrated a new technique to integrate efficient micro-heaters on chip, with an air-cladding resonator, by burying the heater below the optical device layer. Once the buried heater is patterned, encased in SiO$_2$, and planarized, the subsequent process flow is the same as for standard Si$_3$N$_4$ photonic devices. We show that such buried heater is 5$\times$ more efficient than lateral heaters and exhibits low cross-talk which could be even further improved by proper thermal shunt of the chip. We then demonstrate use of the heaters with broadband microcombs, and measure/extract tuning of critical comb parameters for use in optical atomic clocks. While the microcomb measurements are indicative of the use of the buried heater technology for nonlinear integrated photonics applications, we envision many other uses in scenarios for which air cladding are required. This in particular would include microresonator-based sensors~\cite{YuNatRevMethodsPrimers2021} and cavity QED experiments with gas-phase atoms~\cite{OBrienNaturePhoton2009,ChangAppl.Phys.Lett.2020}, where both the wide tuning range and negligible impact on resonator quality factor are important factors.

%  oooooooooo.                 .                        .o.                   oooo   .o8           .   
% `888'   `Y8b              .o8                       .888.                  `888  "888         .o8   
%  888      888  .oooo.   .o888oo  .oooo.            .8"888.     oooo    ooo  888   888oooo.  .o888oo 
%  888      888 `P  )88b    888   `P  )88b          .8' `888.     `88.  .8'   888   d88' `88b   888   
%  888      888  .oP"888    888    .oP"888         .88ooo8888.     `88..8'    888   888   888   888   
%  888     d88' d8(  888    888 . d8(  888        .8'     `888.     `888'     888   888   888   888 . 
% o888bood8P'   `Y888""8o   "888" `Y888""8o      o88o     o8888o     `8'     o888o  `Y8bod8P'   "888" 

\noindent \textbf{\large Data availability} \\
The data that supports the plots within this paper and other findings of this study are available from the corresponding authors upon reasonable request.\\

\noindent \textbf{\large Acknowledgements} \\
The author thanks Nikolai Klimov and Feng Zhou for valuable inputs. The authors acknowledge partial funding support from the DARPA APHI, DARPA SAVaNT, and NIST-on-a-chip programs.\\

%  oooooooooo.   o8o   .o8       oooo   o8o            
% `888'   `Y8b  `"'  "888       `888   `"'            
%  888     888 oooo   888oooo.   888  oooo   .ooooo.  
%  888oooo888' `888   d88' `88b  888  `888  d88' `88b 
%  888    `88b  888   888   888  888   888  888   888 
%  888    .88P  888   888   888  888   888  888   888 
% o888bood8P'  o888o  `Y8bod8P' o888o o888o `Y8bod8P' 
%

\clearpage
%  .oooooo..o                             ooo        ooooo               .   
% d8P'    `Y8                             `88.       .888'             .o8   
% Y88bo.      oooo  oooo  oo.ooooo.        888b     d'888   .oooo.   .o888oo 
%  `"Y8888o.  `888  `888   888' `88b       8 Y88. .P  888  `P  )88b    888   
%      `"Y88b  888   888   888   888       8  `888'   888   .oP"888    888   
% oo     .d8P  888   888   888   888       8    Y     888  d8(  888    888 . 
% 8""88888P'   `V88V"V8P'  888bod8P'      o8o        o888o `Y888""8o   "888" 
%                          888                                               
%                         o888o                                              

\beginsupplement
\section{Experimental Setup}

\begin{figure}[!h]
    \centering
    \includegraphics[width=\linewidth]{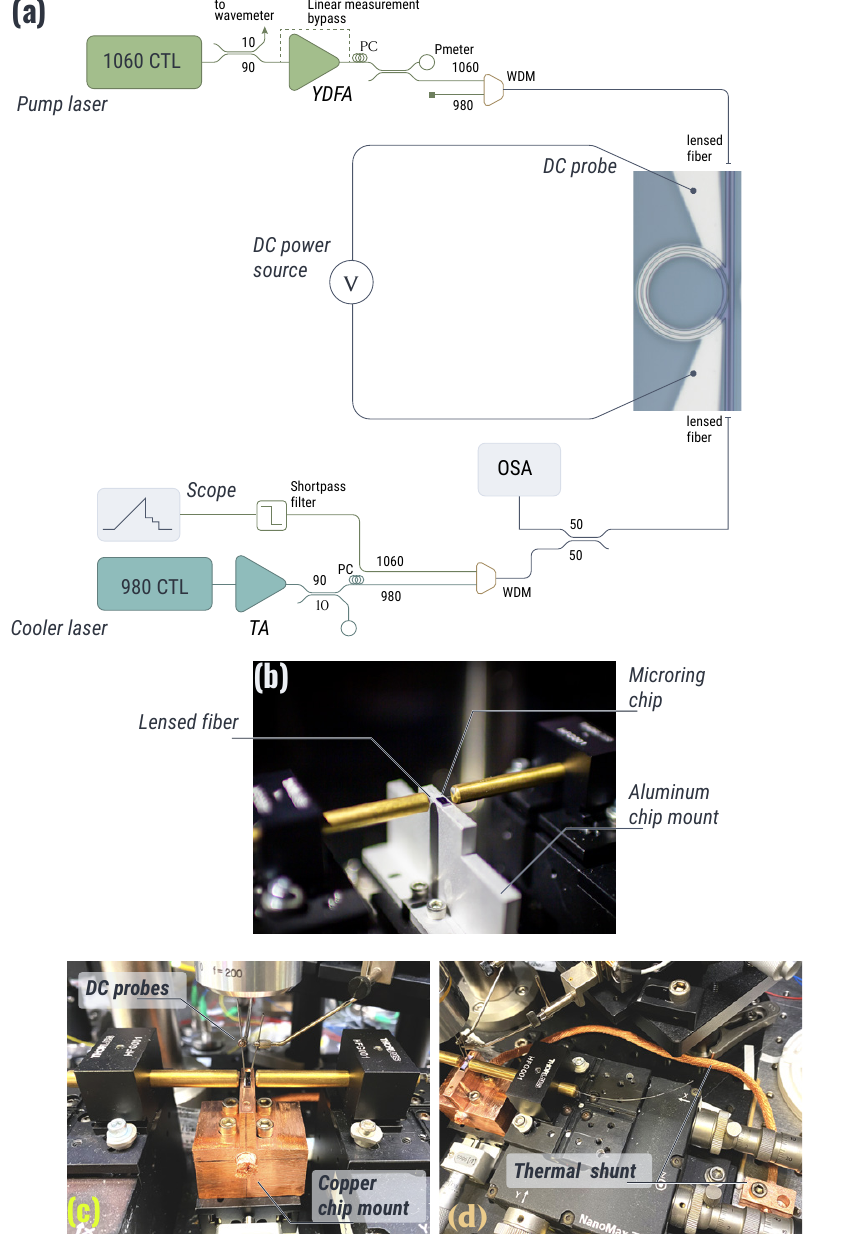}
    \caption{\label{figsup:1}%
    \textbf{Experimental Setup: }%
    \textbf{(a)} Schematic of the experimental setup with a tunable laser around 1060~nm (1060 CTL) for either spectroscopy of cavity modes or to go through a YDFA amplifier for comb generation. In the latter case, an amplified 980~nm tunable laser (980 CTL) is used to actively-cool the resonator allowing for adiabatic DKS access. WDM = wavelength division multiplexer, OSA = optical spectrum analyzer, PC = polarization controller. \textbf{(b)} Close-up picture of the coupling portion of the setup with the aluminum sample holder used in \cref{fig:3}, which acts as a thermal insulator. 
    \textbf{(c)-(d)} Close up of the copper sample mount used in \cref{fig:4}, which is thermally shunted to the optical table and acts as a heat sink at room temperature. This results in a much more stable coupling setup allowing for DKS operation. Note that the coupling was stable enough that the fiber did not need to be adjusted to keep an optimal coupling for any current used in \cref{fig:4}.
    \vspace{5ex}
    }
\end{figure}

\begin{figure}[t]
    \centering
    \includegraphics{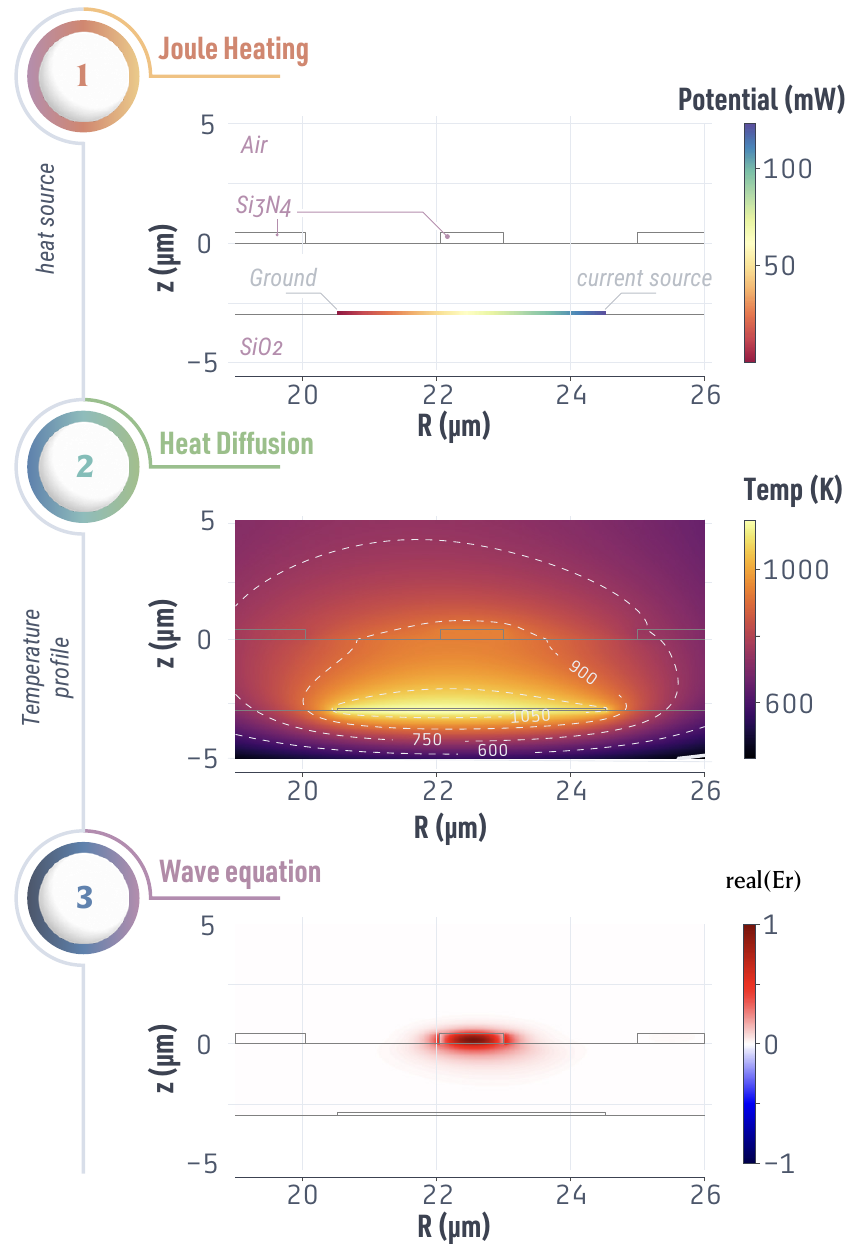}
    \caption{\label{figsup:2}%
    \textbf{Simulation workflow to simulate spectral shifts of the resonant mode of the ring resonator: } 
        (1) we start with a Joule effect simulation in which an input current applied to the buried integrated heater results in an electrical potential and the creation of a heat source. (2) Then, we proceed to solve the Fourier heat equation to find the steady-state of the temperature profile. We apply a fixed boundary condition on the temperature at the bottom of the substrate, and an open boundary elsewhere. We map this temperature profile to the variation of the refractive index profile $\Delta T\partial n / \partial T|_\mathrm{(r,z)}$. (3) This allows for the Maxwell wave equation to account for the temperature profile which will be reflected in the frequency of resonance (i.e. respecting the periodic boundary condition in $\phi$, which is out of the plane of the image). We note that the figures presented are zoomed-in from the full simulation domain.
    }
\end{figure}

The setup used in this work is presented schematically in \cref{figsup:1}(a). The continuously tunable laser (CTL) provides tuning of the wavelength from 1010~nm to 1070~nm to perform spectroscopy, and we adjust the input light polarization such that we excite the the fundamental transverse electric (TE) modes (\textit{i.e.}, TE\textsubscript{0}) of the resonator, by measuring the transmission as a function of wavelength. In this \textit{linear} case, the sample mount is made of aluminum (\cref{figsup:1}(b)), through which we noticed experimentally a limited heat diffusion and an overal increase of the chip temperature, resulting in coupling instability for creation of DKS. In contrast, we use a copper sample mount linked to the optical table(\cref{figsup:1}(c)-(d)) to perform nonlinear measurements, where an ytterbium-doped fiber amplifier (YDFA) is used to produce adequate optical power around 1060~nm, while a (counterpropagating and orthogonally polarized) 980~nm laser is used to actively cool the resonator to reach  states adiabatically. With its improved thermal conductivity and heat sinking to the optical table so that the impact of the buried heaters is more local and the overall chip temperature is not increased as much as in the case of the Al sample mount. Importantly, the overall frequency tuning range achieved by the buried heater is nearly the same regardless of the mount (i.e., comparing \cref{fig:3}(a) and \cref{fig:4}(d)). Interestingly, we believe that use of the copper mount also helps mitigate convective air flows, making the setup much more stable. This has allowed us to measure the DKS states at different electrical powers in \cref{fig:4} without the need to adjust the fiber position for optical coupling.

\section{Thermal Simulations}

In order to accurately account for the integrated buried heater impact on the optical mode of the ring resonator, we performed so called \textit{fully-coupled} simulations, where the Joule heating, heat diffusion and wave equations are solved as a coupled equations system [\cref{figsup:2}]. First, we leverage the symmetry of the system, where in a cylindrical coordinate system $\{r, \phi, z\}$, periodic boundary conditions are assumed in $\phi$, simplifying the system to an effective two-dimensional one along $\{r, z\}$. The optical field is given as $E(r, \phi, z) = E(r, z)e^{i m\phi}$, with $m$ the azimuthal mode number of the ring resonance. Although the implemented heater design breaks this symmetry because it was not fully wrapped around the ring to avoid heating of the resonator-waveguide coupling region, we will assume the heat diffusion is strong enough to allow for uniform temperature in $\phi$. %
Here, we describe the essential steps taken by the solver, although these equations are effectively solved simultaneously. We start by simulating Joule heating, where a current density is injected in the heater assuming the measured resistance of 260~$\Omega$ [\cref{figsup:2}(a)], which we use as a heat power source for the heat diffusion equation [\cref{figsup:2}(b)], allowing us to retrieve the temperature profile along $\{r, z\}$. The boundary conditions are crucial for accurate simulation, and following the improvement of the setup described in the previous section, we assume the bottom of the substrate to be at room temperature and the others as open boundaries. Simulations where the substrate thickness has been increased did not significantly change the simulations results. Although the bottom of the substrate may present a higher temperature than the one assumed here, our assumption of room temperature best reproduces the experimental data presented in Fig.~\ref{fig:4}. Finally, using a Seillmeier model to determine the refractive index of the silicon nitride and silicon dioxide layers according to ellipsometry measurements previously performed~\cite{MoilleOpt.Lett.OL2021}, and accounting for the thermo-refractive index coefficient of each material, we use the simulated profile to determine the refractive index shift $\Delta n(r, z)$ due to the heating. This allows to model the frequency shift of a given resonator mode along with complete dispersion of the resonator modes with electrical power.


\begin{thebibliography}{34}
\bibitem [{NIS()}]{NISTdisclaimer}%
      \BibitemOpen
      \href@noop {} {}\bibinfo {note} {Certain commercial products or names are
      identified to foster understanding. Such identification does not constitute
      recommendation or endorsement by the National Institute of Standards and
      Technology, nor is it intended to imply that the products or names identified
      are necessarily the best available for the purpose.}\BibitemShut {Stop}%
\end{thebibliography}
\end{document}